\documentclass{ws-procs9x6}

\newcommand{\bra}[1]{\langle {#1} |}
\newcommand{\ket}[1]{| {#1} \rangle}

\newcommand{\vecr}{{\mathbf r}}

\newcommand{\veck}{{\mathbf k}}

\begin{document}

\title{Absorbing-boundary-condition method for drip-line nuclei}

\author{T.~Nakatsukasa}

\address{Physics Department, Tohoku University,\\
Sendai 980-8578, Japan}

\author{M.~Ueda and K.~Yabana}

\address{Institute of Physics, University of Tsukuba,\\
Tsukuba 305-8571, Japan}


\maketitle

\abstracts{
Absorbing-boundary-condition method and its applications to nuclear
responses and breakup reactions are reported.
The method facilitates calculations of the continuum states
in the coordinate space of many degrees of freedom.
Properties of nuclei near drip lines are discussed.
}

\section{Absorbing boundary condition (ABC)}
\label{sec:ABC}

Advances in radioactive beams
provide us with a good opportunity to study physics of
weakly-bound finite-quantum systems.
Since excitation spectra above the particle-emission threshold
are continuum spectra,
theoretical analysis requires continuum wave functions.
We have recently investigated an efficient and comprehensive method of
treating the continuum\cite{NY01,NY02-1,NY02-2,NY02-3,YUN02,UYN02}.
This is practically identical to the one called
``Absorbing Boundary Condition (ABC) method''
in the chemical reaction studies\cite{SM92}.
The method allows us to calculate the continuum wave functions
with the outgoing asymptotic behavior of many-body systems in a
finite spatial region where particles interact with each other.

The essential trick for the treatment of the continuum in the ABC method is 
to allow the infinitesimal imaginary part in the Green's function,
$i\epsilon$, to be a function of coordinate and finite, $i\epsilon(\vecr)$.
The $\epsilon(\vecr)$ should be zero in the interacting region and
be positive outside the physically relevant region of space.
Wave functions obtained using the ABC method are meaningful only in the
interacting region.
However, this is enough to solve scattering problems of particles which
interact by a force with finite range.

In order to understand how the ABC method is functioning in later
applications, it is useful to consider the potential scattering of a particle.
The scattering wave function is given by
\begin{equation}
\label{scattering_wave}
\ket{\psi_{\veck}^{(+)}} = \ket{\veck} + \frac{1}{E-H+i\epsilon}V\ket{\veck}
                         \equiv \ket{\veck} + \ket{\psi^{(+)}_{\rm scat}}.
\end{equation}
The scattering amplitude, $f(\Omega)$, is usually
defined by its asymptotic behavior
\begin{equation}
\psi_{\veck}^{(+)}(\vecr) \rightarrow \exp(i\veck\cdot\vecr)
                                   +f(\Omega)\frac{\exp(ikr)}{r} ,
\quad\quad (r\rightarrow\infty),
\end{equation}
but can be written in a form
\begin{equation}
\label{f}
f(\Omega)=-\frac{m}{2\pi\hbar^2}\bra{\veck'}V\ket{\psi_{\veck}^{(+)}}
  =-\frac{m}{2\pi\hbar^2}
      \int d\vecr \exp(-i\veck'\cdot\vecr)V(\vecr)\psi_{\veck}^{(+)}(\vecr) ,
\end{equation}
where $\Omega$ is the direction of $\veck'$ and $|\veck'|=|\veck|$.
Equation (\ref{f}) implies that the $f(\Omega)$ can be
determined by the scattering wave function,
$\psi^{(+)}_\veck(\vecr)$, in the interacting region where $V\neq 0$.
In other words, behavior of $\psi^{(+)}_\veck(\vecr)$ outside the
interacting region is irrelevant to determination of the scattering
properties.

\begin{figure}
\centerline{\resizebox{0.66\textwidth}{!}{%
  \includegraphics{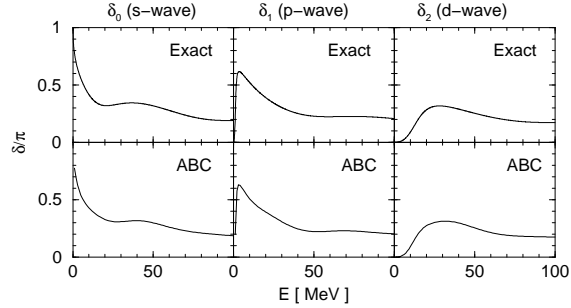}
}}
\caption{Calculated phase shifts for a nucleon scattered by
a square-well potential. See text for explanation.
}
\label{fig:phase_shift}
\end{figure}

Figure~\ref{fig:phase_shift} shows calculated phase shifts for
a nucleon scattered by a square-well potential:
\begin{equation}
V(r)=\cases{ -V_0 & \mbox{for $r<r_0$},\cr
             0    & \mbox{for $r>r_0$},\cr }
\end{equation}
where $V_0=-20$ MeV and $r_0=3$ fm.
The upper part of Fig.~\ref{fig:phase_shift} shows the phase shifts
calculated with a standard numerical procedure.
Namely, we solve the radial Schr\"odinger equations to obtain
the regular wave functions of $s$-, $p$-, and $d$-waves,
and determine the phase shifts at a certain point of $r>r_0$.

On the other hand, Eqs.~(\ref{scattering_wave}) and
(\ref{f}) supply another way of solving the problem.
The $\ket{\psi^{(+)}_{\rm scat}}$ in Eq.~(\ref{scattering_wave})
satisfies the Schr\"odinger equation with a source term
\begin{equation}
\label{Schroedinger_eq_with_source}
(E-H+i\epsilon)\ket{\psi^{(+)}_{\rm scat}} = V\ket{\veck} .
\end{equation}
This equation must be solved with the outgoing boundary condition.
Here, the ABC plays an essential role.
The infinitesimal quantity $i\epsilon$
in Eq.~(\ref{Schroedinger_eq_with_source})
is now replaced by $i\epsilon(\vecr)$,
then, the equation can be solved with the vanishing boundary condition.
The conditions and limitations on $\epsilon(\vecr)$ are discussed in
a number of works\cite{NY01,SM92,Chi91}.
As is emphasized before, the obtained wave function,
$\psi^{(+)}_{\rm scat}(\vecr)$, is correct only in the interacting region.
The scattering amplitude, $f(\Omega)$, is calculated using
Eq.~(\ref{f}), and $\arg(f(\Omega))$ corresponds to the phase shift,
which is displayed in the lower part of Fig.~\ref{fig:phase_shift}.
The results are identical to those of the standard method.
The agreement in results of the two methods clearly indicates that
the continuum in the interacting region ($r<r_0$)
is properly taken into account in the ABC calculation.

\section{Application to nuclear breakup reactions of $^{11}$Be}

The solution of the two-body problem in Sec.~\ref{sec:ABC} was trivial.
The ABC did not show any advantage over the standard techniques.
However, quantum three-body scattering problems are much more difficult,
and one can find usefulness of the ABC.

Let us consider a reaction of a projectile, composed of 
core (C) plus neutron (n), on a target nucleus (T). 
Denoting the projectile-target relative coordinates by ${\bf R}$ and 
the neutron-core relative coordinates by ${\bf r}$, the Hamiltonian 
of this three-body system is expressed as
\begin{equation}
H = -\frac{\hbar^2}{2\mu}\nabla_{\bf R}^2
      -\frac{\hbar^2}{2m} \nabla_{\bf r}^2
      +V_{nC}({\bf r}) + V_{nT}({\bf r}_{nT}) 
      +V_{CT}({\bf R}_{CT})
\end{equation}
where $\mu$ and $m$ are the reduced masses of projectile-target relative 
motion and neutron-core relative motion, respectively. $V_{nC}$,
$V_{nT}$, $V_{CT}$ are the interaction potentials of constituent
particles.

The wave function may be expressed as a 
sum of the Coulomb wave in the incident channel and the scattered wave.
\begin{equation}
\Psi^{(+)}({\bf R},{\bf r})
= \psi^{(+)}({\bf R}) \phi_0({\bf r}) + \Psi_{\rm scat}({\bf R},{\bf r})
\end{equation}
where $\phi_0({\bf r})$ is the ground state of the projectile,
described as a n-C bound state.
The $\Psi_{\rm scat}$ satisfies the
following inhomogeneous equation in the ABC,
\begin{eqnarray}
&& \left\{ E + e_0 + i\epsilon_{nC}(r) + i\epsilon_{PT}(R) -H \right\}
\Psi_{\rm scat}({\bf R},{\bf r})  \nonumber\\
&&=
\left\{ V_{nT}({\bf r}_{nT})+V_{CT}({\bf R}_{CT}) 
- V_C \right\}
\psi^{(+)}({\bf R}) \phi_0({\bf r}),
\label{3Bscat}
\end{eqnarray}
where $V_C$ is the Coulomb distorting potential and $e_0$ is the ground-state
energy of the projectile.
One should note that the right hand side
$\left\{ V_{nT} + V_{CT} - V_C \right\} 
\psi^{(+)}({\bf R}) \phi_0({\bf r})$
is a localized function in space,
being analogous to the fact that
right hand side of Eq.~(\ref{Schroedinger_eq_with_source})
is localized.
Numerical details can be found in our recent paper\cite{UYN02}.
We have studied a deuteron breakup reaction and compared our results
with those of the continuum discretized coupled channel (CDCC)
calculation\cite{UYN02}.
The results agree with those of CDCC.
In this paper, we report the application to a breakup reaction of
$^{11}$Be.

The $^{10}$Be-n potential is taken as Woods-Saxon shape whose depth
is set so as to produce the $2s$ orbital binding energy.
We adopt the optical potential\cite{YUN02} for $^{10}$Be-$^{12}$C
and the Becchetti-Greenlees potential\cite{BG69} for $n-^{12}$C.
The radial region up to 30 fm and 50 fm are used for $R$ and $r$, 
respectively.
The $i\epsilon_{nC}(r)$ and $i\epsilon_{PT}(R)$ is non-zero in the region 
20 fm $< R <$ 30 fm and 25 fm $< r <$ 50 fm.
The n-$^{10}$Be relative angular momenta are included up to $l=3$.

\begin{figure}[ht]
\begin{minipage}[t]{0.35\textwidth}
\includegraphics[width=\textwidth]{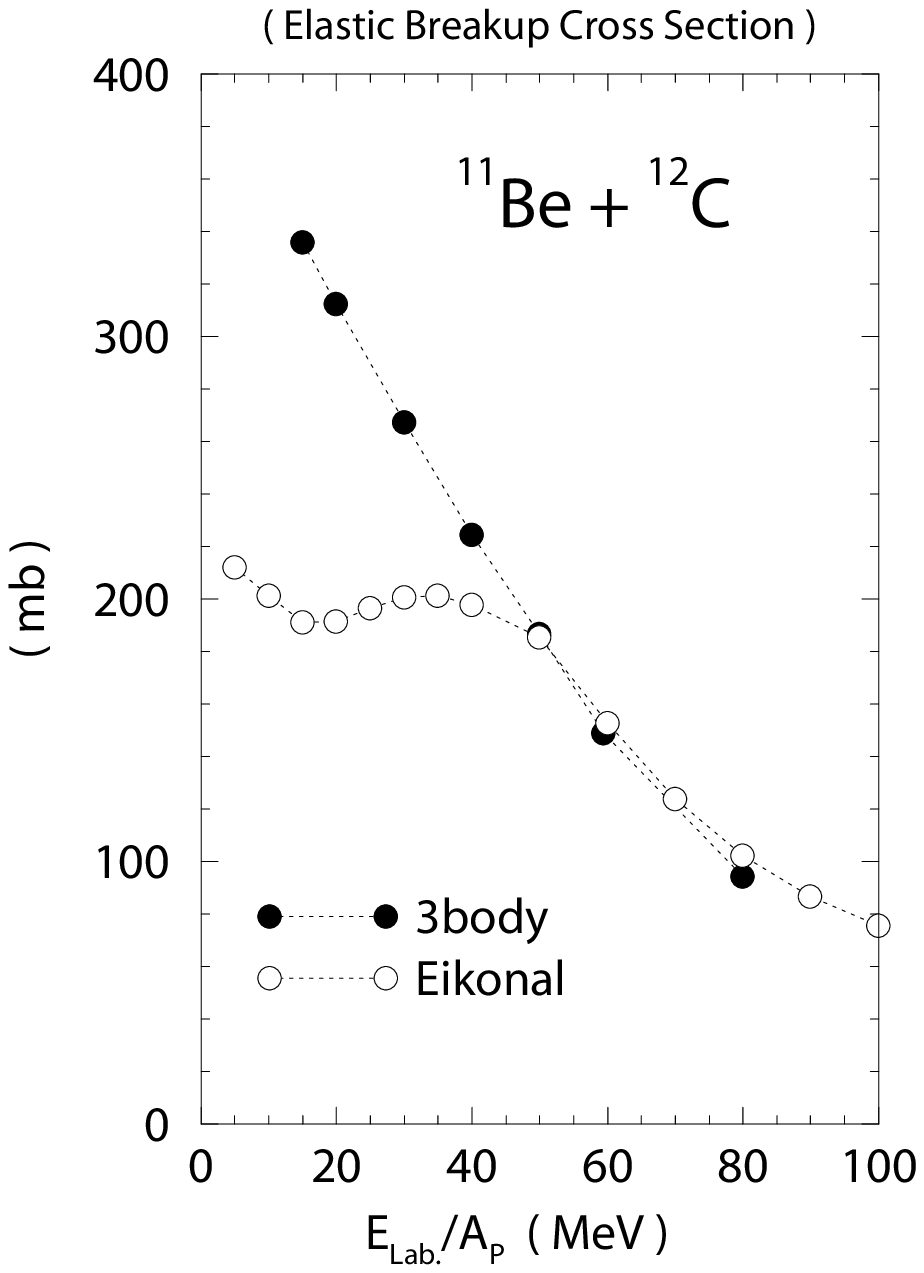}
\caption{
The elastic breakup cross section in $^{11}$Be-$^{12}$C reaction.
Quantum calculation with ABC (closed circles)
is compared with the eikonal calculation (open).
}
\label{fig:11Be}
\end{minipage}
\hfill
\begin{minipage}[t]{0.55\textwidth}
\includegraphics[width=0.8\textwidth]{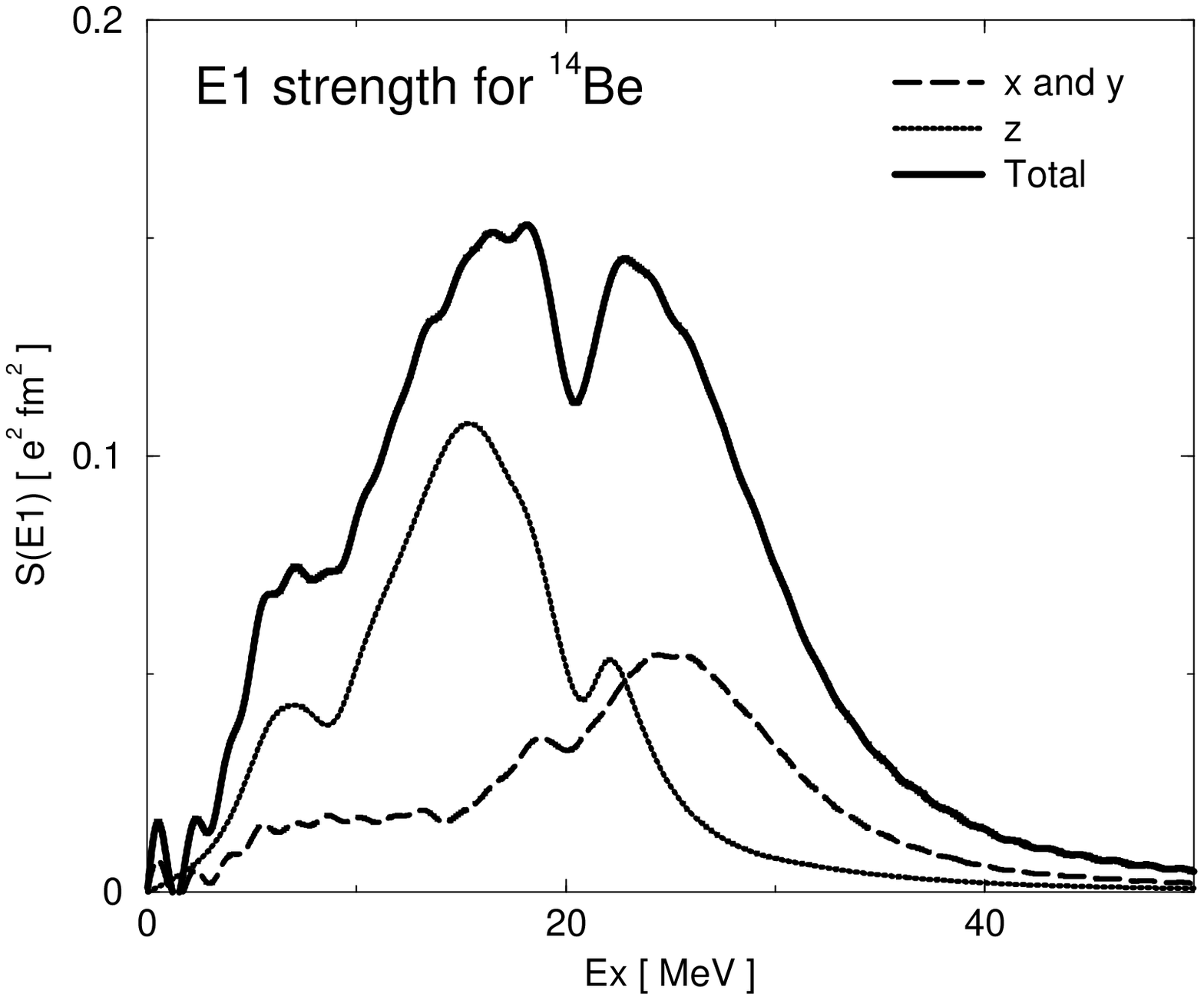}
\caption{Calculated $E1$ strength in $^{14}$Be.
Thin solid and dashed lines indicate the response to dipole fields
parallel and perpendicular to the symmetry axis, respectively.
Thick line shows the total strength.
}
\label{fig:14Be}
\end{minipage}
\end{figure}

In Fig.~\ref{fig:11Be}, we show the elastic breakup cross sections 
of $^{11}$Be-$^{12}$C reaction.
The filled circles are the result of the ABC calculation
and the open circles for the eikonal calculation.
The elastic breakup cross section is substantially larger than that in the 
eikonal approximation at lower incident energy.
The failure of the eikonal approximation is apparent
at the incident energy below 50 MeV/A.
There, the quantum-mechanical treatment is required for the
three-body continuum.

\section{Giant dipole resonances in superdeformed $^{14}$Be}

In studies of giant resonances,
effects of the continuum has been treated in the random-phase
approximation (RPA) with Green's function in the coordinate space\cite{SB75}.
However, it is very difficult to directly apply the method to deformed nuclei
because construction of the Green's function becomes a difficult task
for the multi-dimensional space\cite{NY01}.
We have shown that the ABC method is very useful to treat
the electronic continuum in deformed systems,
such as molecules and clusters\cite{NY01}.
We have also investigated the applicability of the ABC in studies of
nuclear response calculations\cite{NY02-1,NY02-2,NY02-3}.
In this section,
we discuss a giant dipole resonance (GDR) in $^{14}$Be.

We use the ABC in the time-dependent Hartree-Fock (TDHF) calculations
on a three-dimensional (3D) coordinate grid.
In the real-time calculations, the linear response is computed by
applying the isovector-dipole field to the Hartree-Fock (HF) ground state
of $^{14}$Be,
\begin{equation}
{\bf V}_{\rm ext}(t) = k\vecr
  \left\{\frac{1}{2}\left( 1-\tau_z\right)e-\frac{Ze}{A}\right\}
  \delta(t) ,
\label{Vext}
\end{equation}
where $k$ should be small enough to validate
the linear response approximation.
We calculate the expectation values of the $E1$ operator as a function
of time, and then Fourier transforming to get the energy response
\cite{NY01,NY02-1,NY02-2,NY02-3}.
Since all frequencies are contained in the initial perturbation,
the entire energy response can be calculated with a single time evolution.

We use the Skyrme energy functional of EV8 with
the SIII parameter set\cite{BFH87}.
For the time evolution of the TDHF state,
we follow the standard prescription\cite{FKW78}.
The model space is a sphere whose radius is 22 fm.
The $i\epsilon(r)$ is zero in a region of $r<10$ fm,
while it is non-zero at $r>10$ fm.
The TDHF single-particle wave functions
are discretized on a rectangular mesh
in a 3D real space.
Now, we can perform the TDHF simulation together with the
vanishing boundary condition at $r=22$ fm.
Time evolution is carried out up to $T=10\ \hbar$/MeV.

The density distribution of the $^{14}$Be ground state
is calculated to have a prolate superdeformed shape.
Therefore, we expect the deformation splitting of the GDR peak.
The $E1$ oscillator strengths of $^{14}$Be are shown
in Fig.~\ref{fig:14Be}.
Here, we use a smoothing parameter of $\Gamma=1$ MeV.
The calculation predicts the large deformation splitting of about 10 MeV.
However, the width of each peak is so large that the
double-peak structure is almost smeared out in the total
strength (thick line).
This is different from our results of stable nuclei\cite{NY02-1,NY02-3}.
We have observed prominent two-peak structure in those nuclei.
The significant damping width may be a peculiar nature of drip-line nuclei.

\section{Summary}

The idea of absorbing boundary condition is presented.
The continuum is properly treated in the interacting region,
although the equations can be solved with the vanishing boundary condition.
Since we do not need to construct the outgoing boundary wave functions,
the ABC greatly facilitates the scattering problems for cases of
the continuum in many degrees of freedom.
We apply the ABC to the quantum reaction studies of
elastic breakup of $^{11}$Be and
the linear response calculation for
the isovector GDR of superdeformed $^{14}$Be.

This work is supported by
the Grand-in-Aid for Scientific Research (No. 14740146)
from the Japan Society for the Promotion of Science.

\end{document}